\documentclass[aps,twocolumn]{revtex4}
\usepackage{graphics}
\begin{document}
\newcommand{\bstfile}{osa} %alternative styles: osa, prasty or revtex
\newcommand{\bibs}{d:/Dad/Bibliography/TexBiB/final}
\title{Intrinsic hyperpolarizabilities as a figure of merit for electro-optic molecules}
\author{Juefei Zhou and Mark G. Kuzyk}
\affiliation{Department of Physics and Astronomy, Washington State University, Pullman, Washington
99164-2814}
\date{\today}

\begin{abstract}
We propose the scale-invariant intrinsic hyperpolarizability as a measure of the figure of merit for electro-optic molecules.  By applying our analysis to the best second-order nonlinear optical molecules that are made using present paradigms, we conclude that it should be possible to make dye-doped polymers with electro-optic coefficients of several thousand pico-meters per volt.
\end{abstract}

\maketitle

\section{Introduction}

The nonlinear-optical susceptibility of a material determines its suitability in applications such as telecommunications,\cite{wang04.01} three-dimensional nano-photolithography,\cite{cumps99.01,kawat01.01} and new materials development\cite{karot04.01} for novel cancer therapies.\cite{roy03.01}  In the present work, we focus on the molecular hyperpolarizability and how it impacts the efficiency of polymer electro-optic modulators and switches.

Polymeric materials have many useful properties such as the ability to be formed into thin films\cite{pante88.01} and fibers\cite{garve96.01} with electrodes.\cite{welker98.01} Electrooptic modulator devices that are used in fiber-optic system require pigtails that can be connected to a fiber.  Thus, it is necessary to define waveguides in the polymer to form the device.  Thackara and coworkers demonstrated that waveguides could be defined in a dye-doped polymer by elevating the refractive index through the application of an electric field.\cite{thack88.01} In that work, the refractive index increase was due to electric-field-induced alignment of Disperse Red \#1 (DR1) azo dye molecules in the PMMA host polymer.  This technique was technologically attractive because the same electrodes that were used to apply the electric field to define the waveguide were also used to apply the modulating voltage.  Since metal patterning techniques were mature even at that time, making waveguides with electric-field poling provided a straightforward method for making complex structures.

Yang and coworkers showed that polymers could be selectively poled using an electron beam;\cite{yang94.01} and, Shi and coworkers demonstrated that a large refractive index change could be induced in a side chain polymer due to photoisomerization.\cite{shi90.01} This allowed photo-patterning as a means to defining waveguides in a dye-doped polymer with a large electro-optic response.  Furthermore, the side-chain polymers used for making the materials were more stable to photodegradation.\cite{singe88.01,becke94.01}

The first model of electric-field poling for the second-order susceptibility of a poled dye-doped polymer was introduced more than twenty years ago.\cite{singe87.01} This model assumed a dilute collection of non-interacting molecules that rotate freely above the glass transition temperature of the polymer, an assumption that was found to be consistent with experimental evidence.\cite{ghebr95.04,ghebr97.01} At high concentrations, the thermodynamic model for poling predicted exceptionally large electro-optic coefficients that greatly exceeded the performance of lithium niobate.  Harper and coworkers showed that this model overestimated the bulk response because it did not account for chromophore-chromophore electrostatic interactions that become dominant at high concentrations.\cite{harpe98.01}

Ma and coworkers found that incorporating nonlinear chromophores into a dendrimer allowed the material to be efficiently poled with an electric field.\cite{ma01.01} The material yielded a large electrooptic coefficient ($60 \, pm/V$), and was stable to orientational decay.  Sullivan and coworkers demonstrated a tri-component poled polymer made with a dendrimer containing (1) an electrooptic chromophore, (2) a second electro-optic chromophore that acts as a cross linker, and (3) a third optically inert chromophore that controls the glass transition temperature of the composite system.\cite{sulli07.01}  The electrooptic coefficient of $150 \, pm/V$ was found to be stable at $130^o \,C$ over 15 months.  Such properties make these materials ideal for electrooptic devices.  Indeed, Michalak et al made several micro-channel $50 \, GHz$ modulators with $V_{\pi} < 12 \, V$ on a wafer using amorphous polycarbonate doped with an AJL8 chromophore;\cite{Micha06.01} Shi et al made a sub-1-volt modulator operating at $1318 \, nm$ by controlling the chromophore shape to optimize the effects of electric field poling-induced electro-optic activity by reducing intermolecular
electrostatic interactions;\cite{shi00.01} Lee and coworkers demonstrated that a polymer modulator could modulate signal up to a bandwidth of $200 \, GHz$;\cite{lee02.01} and, Rabiei and coworkers demonstrated polymeric micro-ring filters and modulators.\cite{rebie02.01}

Combinations of various approaches to arrange chromophores with polar order have lead to efficient and stable high-speed modulators.  However, the fundamental limit of the efficiency of such devices is ultimately constrained by the limit of the nonlinear response of the chromophores.  In this work, we analyze the best device materials to gain an understanding of the amount of future improvements that are attainable by invoking the fundamental limits of the nonlinear-optical response of the hyperpolarizability.\cite{kuzyk00.01,kuzyk03.02}  Until a recent breakthrough in molecular design using modulation of conjugation (MOC),\cite{perez07.01,zhou06.01,zhou07.02} the best molecules fell a factor of 30 short of the fundamental limit.\cite{Tripa04.01}

\section{Theory}

The bulk second-order nonlinear optical response originates in the hyperpolarizability, $\beta$, of the molecules from which the material is made.  When assessing molecules for applications in electrooptic switching/modulation or frequency doubling, it is natural to compare their hyperpolarizabilities.  However, such comparisons are complicated by the fact that the hyperpolarizability depends on wavelength.  Even when two molecules are measured at the same wavelength, one of them could have a resonantly enhanced hyperpolarizability making its response appear larger.  To take into account resonance enhancement, it has become a common practice to use a two-level model\cite{oudar77.03} to extrapolate the measured values of the hyperpolarizability to the off-resonant regime.

The off-resonant hyperpolarizability, $\beta_0$ is related by the two-level model to the second harmonic hyperpolarizability, $\beta$, measured with a fundamental photon energy $\hbar \omega$  according to,\cite{ikeda91.01}
\begin{equation}\label{beta0}
\beta_0 = \beta \cdot \left( 1 - \left(\frac{\hbar \omega} {E_{10}}\right)^2 \right) \left( 1 - 4\left(\frac{\hbar \omega} {E_{10}}\right)^2 \right),
\end{equation}
where $E_{10} = E_1 - E_0$ is the energy difference between the first excited state and the ground state.  This quantity is an attempt to define a fundamental property of a molecule that can be compared with others.

Until recently, there were two ways to evaluate a molecule's nonlinear response. Molecular-engineering studies focused on $\beta_0$ as a frequency-independent measure to assess what was believed to be an intrinsic molecular property. For a particular device application, on the other hand, researchers directly evaluated the nonlinear response of the material at the wavelength of device operation.

Recently, it has been proposed that a better measure of the intrinsic hyperpolarizability of a molecule is the ratio of the off-resonant hyperpolarizability to the off-resonant fundamental limit of the hyperpolarizability,\cite{zhou06.01} \begin{equation}\label{betaintrinsic}
\beta_0^{int} = \beta_0/\beta_0^{MAX},
\end{equation}
where the fundamental limit is given by,
\begin{equation}\label{betalimit}
\beta_{0}^{MAX} = \sqrt[4]{3} \left( \frac {e \hbar} {\sqrt{m}} \right)^3 \frac {N^{3/2}} {E_{10}^{7/2}},
\end{equation}
where $N$ is the number of electrons in the system, $m$ the electron mass, $e$ the electron charge and $\hbar$ is Planck's constant.

The significance of the intrinsic hyperpolarizability can be better understood by analyzing the scaling properties of the one-dimensional Schrodinger Equation,
\begin{equation}\label{schrodinger}
- \frac {\hbar^2} {2m} \frac {\partial^2} {\partial x^2} \psi(x) + V(x) \psi(x) = E \psi(x) ,
\end{equation}
where the potential energy function $V(x)$ quantifies the forces experienced by the electrons in a molecule, and depends on the positions and charges of the nuclei.  If we re-scale the Schrodinger equation by taking $x \rightarrow \epsilon y$, where $\epsilon$ is a scaling factor, we get,
\begin{equation}\label{schrodinger2}
- \frac {\hbar^2} {2m} \frac {\partial^2} {\partial y^2} \psi(\epsilon y) + \epsilon^2 V(\epsilon y) \psi(\epsilon y) = \epsilon^2 E \psi(\epsilon y) .
\end{equation}
Thus, if $\psi(x)$ is a solution of the Schrodinger Equation for potential $V(x)$ with eigen energy $E$, $\psi(\epsilon x)$ will be a solution of the Schrodinger Equation for the potential $\epsilon^2 V(\epsilon x)$ with eigen energy $\epsilon^2 E$.  The matrix elements of the position operator, $x_{ij}$ are defined by:
\begin{equation}\label{positionMatrix}
x_{ij} = \frac {\int_{-\infty}^{+\infty} \psi_i^* (x) x \psi_j (x) \, dx} {\int_{-\infty}^{+\infty} \psi_i^* (x) \psi_j (x) \, dx},
\end{equation}
which upon re-scaling the wavefunctions by $\epsilon$ becomes,
\begin{equation}\label{positionMatrix2}
x_{ij}' = \frac {\int_{-\infty}^{+\infty} \psi_i^* (\epsilon x) x \psi_j (\epsilon x) \, dx} {\int_{-\infty}^{+\infty} \psi_i^* (\epsilon x) \psi_j (\epsilon x) \, dx} = \frac {x_{ij}} {\epsilon}.
\end{equation}

The hyperpolarizability is of the form,\cite{orr71.01}
\begin{eqnarray}\label{beta}
\beta_{xxx} (\omega_1, \omega_2) & = & -   \frac {e^3} {2}
P_{\omega_1, \omega_2} \left[ {\sum_{n}^{\infty}} ' \frac {\left|
x_{0n} \right|^2 \Delta x_{n0}} {D_{nn}^{-1}(\omega_1,\omega_2)}
\right. \\ \nonumber & + & \left.{\sum_{n }^{\infty}} ' {\sum_{m
\neq n }^{\infty}} ' \frac {x_{0n} x_{nm} x_{m0}}
{D_{nm}^{-1}(\omega_1,\omega_2) }\right],
\end{eqnarray}
where $-e$ is the electron charge,
$\Delta x_{n0} = x_{nn} - x_{00}$ is the difference in the
expectation value of the electron position between state $n$ and the
ground state.  The primes indicate that the ground
state is excluded from the sum and the permutation operator
$P_{\omega_1, \omega_2}$ directs us to sum over all six frequency
permutations.  $D_{nm}^{-1}(\omega_1,\omega_2)$ gives the dispersion
of $\beta$ and $\hbar \omega_1$ and $\hbar \omega_2$
are the photon frequencies,
\begin{eqnarray}\label{beta denominators}
&& P_{\omega_{1}, \omega_{2}} [D_{nm} (\omega_1 , \omega_2)]  =  \frac {1} {2 \hbar^2} \left[ \frac {1} { \left(\omega_{n0} - \omega_1 - \omega_2 \right) \left(\omega_{m0} - \omega_1 \right)} \right. \nonumber \\
& + &  \frac {1} {\left(\omega_{n0}^* + \omega_2 \right) \left(\omega_{m0} - \omega_1 \right)}  \nonumber \\
& + &  \frac {1} {\left(\omega_{n0}^* + \omega_2 \right) \left(\omega_{m0}^* + \omega_1 + \omega_2 \right)} \nonumber \\
& + & \left. \omega_1 \leftrightarrow \omega_2 \hspace{1em}
\mbox{for the three previous terms} \right],
\end{eqnarray}
and where $\omega_{m0} = \omega_{m0}^0 - i \gamma_{m0}$.  $\hbar
\omega_{m0}^0$ is the energy difference between state $m$ and the
ground state and $\gamma_{m0}$ is half the natural linewidth for a
transition from state $m$ to the ground state.

According to Equations \ref{beta} and \ref{beta denominators}, the hyperpolarizability is proportional to position matrix elements to the third power and inversely proportional to the energy squared.  According to Equation \ref{schrodinger2} and Equation \ref{positionMatrix2}, the off-resonance hyperpolarizability (i.e. in the zero-frequency limit) scales as
\begin{equation}\label{BetaScale}
\beta \propto \frac {x^3} {E^2} \rightarrow \frac {(x/\epsilon)^3} {(\epsilon^2 E)^2} = \frac {x^3} {\epsilon^7 E^2} ,
\end{equation}
where $E$ and $x$ represent energy eigenvalues and position matrix elements.  But, the intrinsic hyperpolarizability scales according to
\begin{equation}\label{BetaIntScale}
\beta_0^{int} = \frac {\beta_0} {\beta_0^{MAX}} \propto \frac {x^3} {E^2} \cdot \frac {E_{10}^{7/2}} {N^{3/2}} \rightarrow \frac {(x/\epsilon)^3} {(\epsilon^2 E)^2} \cdot \frac {(\epsilon^2 E_{10})^{7/2}} {N^{3/2}} = \frac {x^3} {E^2} .
\end{equation}
Because Equation \ref{BetaIntScale} clearly shows that $\beta_0^{int}$ is independent of $\epsilon$, it is a scale invariant quantity.  Similarly, the factor $N^{3/2}$ normalizes for the the number of electrons.  Thus, the intrinsic hyperpolarizability removes size effects and allows molecules of drastically differing sizes to be compared directly.  So, while a large molecule may have a much larger zero-frequency hyperpolarizability than a smaller one, the smaller molecule may be intrinsically more efficient when size effects are taken into account.

The assumption that $\beta_0$ alone is a reasonable measure of a molecule's off-resonance response is flawed because of two serious problems.  First, the two-level dispersion model may be highly inaccurate.  Indeed, the excited state transition moments and energies for many states are often needed to accurately calculate the dispersion of the hyperpolarizability\cite{Dirk89.01,champ06.01} and there are well-known problems associated with extrapolating $\beta_0$ to zero frequency.\cite{kaatz98.01,woodf99.01}  Even when only two electronic states dominate the response, the damping correction must be carefully taken into account\cite{dirk90.01,berko00.01} and vibronic overtones may complicate the analysis.\cite{wang00.01,wang01.01} Secondly, there is nothing intrinsic about $\beta_0$ as we can argue from the fact that it can be large for very different underlying reasons.  For example, $\beta_0$, can be large when the transition moment to the lowest-energy excited state is large; or, when the transition moment is small but the first excited state energy is low due to resonance enhancement of the zero-frequency response.  The intrinsic hyperpolarizability defined by Equation \ref{betalimit} with scaling given by Equation \ref{BetaIntScale} removes this resonance effect.

Clearly, $\beta_0$ is not a good metric for comparing molecules if one's goal is to determine the origin of the nonlinear-optical response; nor is it a good metric for evaluating materials for their usefulness in a particular device.  When studying the  fundamental properties of materials, or assessing the suitability of a material for an application, it would be best to have an absolute standard at the appropriate set of wavelengths rather than relying on a two-level extrapolation of the measured hyperpolarizability to get $\beta_0$.  We note that $\beta_0$ is also sometimes called the intrinsic hyperpolarizability,\cite{wang01.01} which it is not.  So, while the use of $\beta_0$ for comparing molecules is a common practice, we argue that that it does not elucidate any fundamental property of a molecule, nor is it useful for for predicting the response at an arbitrary wavelength.  We propose that a better measure of the intrinsic response is the ratio of the measured value of the hyperpolarizability to the fundamental limit at that set of wavelengths,\cite{kuzyk06.03}
\begin{equation}\label{newIntriniscBeta}
\beta^{int} (\omega_1 , \omega_2) = \frac {\beta(\omega_1 , \omega_2)} {\beta^{MAX} (\omega_1 , \omega_2)},
\end{equation}
where\cite{kuzyk06.03}
\begin{equation}
\beta_{xxx}^{MAX} (\omega_1, \omega_2) = \beta_{0}^{MAX} \cdot \frac {1} {6} \frac {E E_{10}^2} {\sqrt{1-E}} \cdot D^{3L} (\omega_1, \omega_2) ,
\label{beta contracted-Dispersion}
\end{equation}
and where
\begin{eqnarray}\label{beta contracted-threelevel}
D^{3L} (\omega_1, \omega_2) & = & \left[
\frac {1} {D_{12}^{-1} (\omega_1 , \omega_2)} -  \frac { \left( 2
\frac {E_{20}} {E_{10}} -1 \right) } {D_{11}^{-1} (\omega_1 ,
\omega_2)} \right. \\ \nonumber & + & \left. \frac {1} {D_{21}^{-1}
(\omega_1 , \omega_2)} - \frac { \left( 2 \frac {E_{10}} {E_{20}} -1
\right) } {D_{22}^{-1} (\omega_1 , \omega_2)} \right].
\end{eqnarray}
Note that the dipole-free SOS expression was used to determine the above results.\cite{kuzyk05.02}  This new generalized intrinsic hyperpolarizability provides an absolute comparison between the measured or calculated response of a molecule with an absolute standard of the nonlinear response at any set of wavelengths. Thus, the generalized intrinsic hyperpolarizability, $\beta^{int} (\omega_1 , \omega_2)$ provides a figure of merit for any nonlinear-optical process at any set of wavelengths, a more powerful analytical tool than a single wavelength-independent number such as $\beta_0$.  Note that $\beta^{int} (\omega_1 , \omega_2) \leq 1$

When calculating the wavelength dependence of the fundamental limits using Equation \ref{beta contracted-threelevel} near resonance, a Lorentzian damping factor is used.\cite{kuzyk06.03}  Since real systems are inhomogeneously broadened, Lorentzian broadening must be augmented with statistical averaging of the excited state energies.  Studies of broadening suggest that it is important to use appropriate models when analyzing the dispersion of both the linear and nonlinear response.\cite{kruhl08.01,kruhl08.02}

\section{Discussion}

\begin{widetext}

\begin{table}\caption{Molecular data.\label{tab:data}}
\begin{tabular}{c c c c c c c c c c c c c c c}
  \hline
  Molecule & $\beta$ & & $E_{10}$ & $E_{20}$ & $\Gamma_1$ & $\Gamma_2$ & $N$ & $\hbar \omega$ & & $\beta_0$ & $\beta_0^{MAX}$ & & $\beta_0^{int}$ & $\beta^{int}(\omega,\omega)$\\
 $[ref]$ & (esu) & & (eV) & (eV) & (eV) & (eV) & & (eV) & & (esu) & (esu) & & &\\
  \hline
    1\cite{liao05.01} & $7.36(\pm 0.81) \times 10^{-29}$ & & 1.98 & 2.95 & 0.20 & 0.26 & 22 & 1.24 & & $2.55 \times 10^{-29}$ & $1.12 \times 10^{-26}$ & & 0.002 & 0.007 \\
    2\cite{liao05.01} & $1.34(\pm 0.54) \times 10^{-28}$ & & 1.80 & 2.80 & 0.20 & 0.27 & 24 & 1.24 & & $6.33 \times 10^{-29}$ & $1.78 \times 10^{-26}$ & & 0.004 & 0.013 \\
    3\cite{liao05.01} & $7.7(\pm 1.7) \times 10^{-28}$ & & 1.80 & 2.40 & 0.16 & 0.25 & 28 & 1.24 & & $3.64 \times 10^{-28}$ & $2.25 \times 10^{-26}$ & & 0.016 & 0.165 \\
    4\cite{liao05.01} & $9.7(\pm 1.4) \times 10^{-28}$ & & 1.60 & 2.20 & 0.16 & 0.25 & 30 & 1.24 & & $5.44 \times 10^{-28}$ & $3.76 \times 10^{-26}$ & & 0.014 & 0.020 \\
    TM-2\cite{Kang07.01} & $8.4(\pm 1.2) \times 10^{-27}$ & & 2.18 & - & - & - & 20 & 0.65 & & $4.96 \times 10^{-27}$ & $6.93 \times 10^{-27}$ & & 0.715 & - \\
    TMC-2\cite{Kang07.01} & $8.9(\pm 1.3) \times 10^{-28}$ & & 2.18 & 3.95 & 0.25 & 0.25 & 16 & 0.65 & & $5.21 \times 10^{-28}$ & $4.96 \times 10^{-27}$ & & 0.105 & 0.169 \\
    TMC-3\cite{Kang07.01} & $9.6(\pm 1.5) \times 10^{-27}$ & & 2.30 & 2.86 & 0.25 & 0.25 & 24 & 0.65 & & $6.04 \times 10^{-27}$ & $7.56 \times 10^{-27}$ & & 0.799 & 2.96 \\
   \hline
\end{tabular}
\end{table}

\end{widetext}

\begin{figure}
\includegraphics{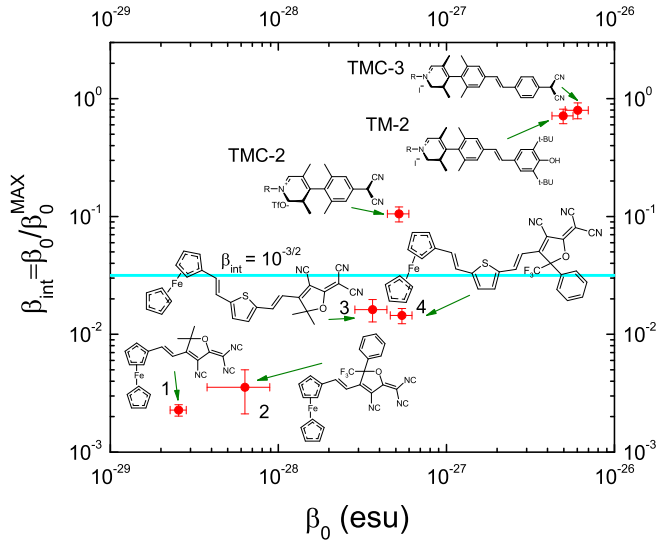}
\caption{Intrinsic off-resonant hyperpolarizability, $\beta_0/\beta_0^{MAX}$, as a function of off-resonant hyperpolarizability, $\beta_0$.}
\label{figure1}
\end{figure}

Table \ref{tab:data} summarizes our analysis of experimental results found in the literature\cite{liao05.01,Kang07.01} and Figure \ref{figure1} shows a plot of the intrinsic off-resonant hyperpolarizability as a function of the measured off-resonant hyperpolarizability.  The molecular structures in Figure \ref{figure1} appear beside the data points and are labeled 1,2,3,4,TM-2, TMC-2, and TMC-3 in correspondence to the labels in the table.  Molecules TM-2, TMC-2, and TMC-3 are measured using liquid solution electric field-induced second harmonic (EFISH) generation while molecules 1-4 are measured using hyper-Rayleigh scattering (HRS).  The horizontal line at $\beta_0^{int} = 10^{-3/2}$ represents the best molecules reported prior to 2007.\cite{Tripa04.01}  Only recently have molecules been synthesized that break through this limit.\cite{perez07.02,perez07.01}

The intrinsic hyperpolarizability is a function of the number of conjugated electrons and energy difference between ground and excited state.  If the conjugations path is broken, the system behaves approximately as independent molecules - in which case the hyperpolarizability is simply the sum over the hyperpolarizabilities of the various independent parts of the molecule.  Thus, if a molecule is made of $n$ independent subunits, and subunit $i$ has $N_i$ electrons and an energy difference between ground and first dominant excited state $E_{10}^i$, Equation \ref{betalimit} can be re-expressed as
\begin{equation}\label{betalimitsum}
\beta_{0}^{MAX} = \sum_i^n \sqrt[4]{3} \left( \frac {e \hbar} {\sqrt{m}} \right)^3 \frac {N_i^{3/2}} {(E_{10}^i)^{7/2}}.
\end{equation}

The energies of the various parts of a molecule are not generally known {\em a priori}, so it is not a simple matter to perform the sum in Equation \ref{betalimitsum}.  Usually, the lowest energy state of a molecule corresponds to the excitation of the largest subunit, and, the largest subunit is usually much larger than any of the others.  Calling the lowest energy state $E_{10}$, we can reexpress the fundamental limit in the form:
\begin{eqnarray}\label{betalimitapprox}
\beta_{0}^{MAX} & = & \sqrt[4]{3} \left( \frac {e \hbar} {\sqrt{m}} \right)^3 \frac {1} {E_{10}} \sum_i^n N_i^{3/2} \left(  \frac {E_{10}} {(E_{10}^i)} \right)^{7/2} \nonumber \\
& \leq & \sqrt[4]{3} \left( \frac {e \hbar} {\sqrt{m}} \right)^3 \frac {1} {E_{10}} \sum_i^n N_i^{3/2} \nonumber \\
& \equiv & \sqrt[4]{3} \left( \frac {e \hbar} {\sqrt{m}} \right)^3 \frac {1} {E_{10}} N^{3/2}
\end{eqnarray}
where $N$ is the effective number of electrons given by
\begin{equation}\label{countelectrons}
N = \left(\sum_i^n N_i^{3/2} \right)^{2/3}.
\end{equation}
When all of the parts of the molecule are of approximately the same size, Equation \ref{countelectrons} yields a good approximation of the number of electrons and therefore a good estimate of the fundamental limit.  If the largest subunit is much larger than the others, Equation \ref{countelectrons} overestimates the number of electrons and thus leads to an overestimate of the fundamental limit.  In this case, counting only the electrons in the largest part of the molecule may give the best estimate.  The number of electrons listed for each molecule in Table \ref{tab:data} is calculated using Equation \ref{countelectrons}.

Molecules 1-4 all fall below $\beta_0^{int} = 10^{-3/2}$, the longstanding and unexplained ceiling of all measured molecules prior to 2007.\cite{Tripa04.01}  Molecules 3 and 4 have an intrinsic hyperpolarizability that is as large as some of the best molecules ever synthesized.  While Molecule 4 has a larger value of $\beta_0$ than Molecule 3, its intrinsic hyperpolarizability is smaller than Molecule 3's.  When designing molecules for a particular off-resonant application, using the value of $\beta_0$ as a figure of merit may be appropriate for order-of-magnitude comparisons.  However, since comparisons of the intrinsic hyperpolarizability removes the effects of scaling, it makes clear that Molecule 4 is not intrinsically as good as Molecule 3.  Thus, in order to take full advantage of all of the electrons in the molecule, only a {\em scalable} paradigm that uses the structure of Molecule 3 as a basis would be promising.  Indeed, a similar analysis of the two-photon absorption cross-sections\cite{perez05.01} shows that while longer molecules may have a larger two-photon cross-section, the intrinsic two-photon cross-sections of the longer ones actually decrease with length - illustrating how a paradigm can be non-scalable.\cite{kuzyk03.03}  This shows how the intrinsic optical properties of a molecule better quantify a molecule's strength of interaction with light, which can be overlooked when only quantities such as $\beta_0$ are compared.

The twisted molecules TM-2 and TMC-3 have the largest values of $\beta_0$ and their intrinsic off-resonant hyperpolarizabilities are right at the fundamental limit.  They are the first nonlinear-optical chromophores that fall substantially above $\beta_0^{int} = 10^{-3/2}$.  It is useful to compare similar molecules of differing size to determine whether this molecular paradigm is scalable.  Molecule TMC-3 is made by increasing the length of the bridge in Molecule TMC-2.  $\beta_0$ of Molecule TMC-3 is about an order of magnitude larger than Molecule TMC-2; but most importantly, the intrinsic hyperpolarizability is also larger by almost an order of magnitude.  Twisted chromophores\cite{Kang07.01} are thus an example of a new scalable paradigm where larger molecules in the series take full advantage of the additional electrons.

Previous calculations that used numerical optimization to identify molecular design rules found that the best molecules have oscillating potential energy functions, which led Zhou and coworkers to suggest that modulation of conjugation may lead to improved nonlinear response.\cite{zhou06.01} Perhaps the twist in the molecule provides an effective kink in the potential energy function that leads to an enhancement.
\begin{figure}
\includegraphics{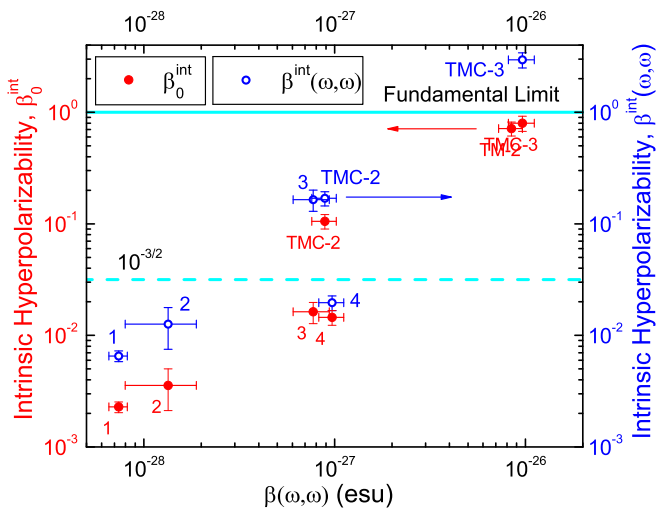}
\caption{Intrinsic hyperpolarizability at the measurement wavelength, $\beta^{int}(\omega,\omega)$, and off resonance, $\beta_0/\beta_0^{MAX}$, as a function of the measured hyperpolarizability, $\beta(\omega,\omega)$.}
\label{figure2}
\end{figure}
Figure \ref{figure2} shows a plot of the intrinsic hyperpolarizability at the measurement wavelength, $\beta^{int}(\omega,\omega)$, and the off resonance value, $\beta_0/\beta_0^{MAX}$, as a function of the measured hyperpolarizability, $\beta(\omega,\omega)$.  The intrinsic hyperpolarizability at the measurement wavelength, $\beta(\omega,\omega)$, is systematically larger than $\beta_0^{int}$.  We stress that both of these metrics of the intrinsic hyperpolarizability must be smaller than unity.  Since Molecules 1-4 were measured at an incident photon energy twice that of the measurements on the twisted systems, each group needs to be considered separately since the intrinsic hyperpolarizability is being determined at a different fundamental wavelength.

Molecule 3 shows the largest difference between $\beta_0^{int}$ and $\beta^{int}(\omega, \omega)$, with $\beta^{int}(\omega, \omega)$ substantially above $10^{-3/2}$. This is not surprising because the measurement is on resonance with two-photon excitation, that is, $2 \hbar \omega \approx E_{20}$.  However, being near resonance is no guarantee of having a large intrinsic hyperpolarizability.  For example, Molecule 4, has the same low intrinsic hyperpolarizability on and off resonance of about $0.02$.  This analysis clearly shows how two similar molecules may have very different intrinsic hyperpolarizabilities near resonance even when they are comparable off resonance.  In principle, the wavelength-dependent intrinsic hyperpolarizability, $\beta^{int}(\omega,\omega)$, takes into account resonant enhancement at any wavelength, and should therefore be the best indicator of a molecule's nonlinear-optical response.

The twisted systems, on the other hand, all have large intrinsic hyperpolarizability $\beta_0^{int}$, and the resonant intrinsic hyperpolarizabilities are even larger.  In fact, Molecule TMC-3 is above the fundamental limit by a factor of 3.  So, either the theory is flawed or the measurements are inaccurate.  Since nonlinear-optical measurements are inherently complex, there are several possible explanations for getting anomalously high experimental results.  To get vacuum susceptibilities from liquid solution measurements, it is typical for researchers to use Lorentz local field models.  Wortmann and Bishop have shown that such models can be off by factors of two.\cite{wortm98.01}  It is plausible that such issues along with experimental uncertainties could have lead to an un-physically high value of the intrinsic nonlinear-optical response.  Perhaps inhomogenous broadening needs to be included in the theory;\cite{kruhl08.01,kruhl08.02} or, the conventions of expressing $\beta$ are are not consistent. Alternatively, the theory may need to be revisited for the possibility that the fundamental limits may be underestimated.

Whatever the resolution, the twisted nonlinear-optical chromophores show the potential for further improvements in the hyperpolarizability that can be subsequently made into bulk materials with exceptional nonlinear response.  A comparison of the twisted molecules with present molecules leads to a better appreciation for the potential impact of such systems.  The twisted molecular systems have an intrinsic nonlinear-optical response that is a factor of 30 better than all others.  Present state-of-the-art molecules, when made into a poled polymer, have electro-optical coefficients of $169 \, pm/V$ and are highly thermally stable.\cite{cheng07.01} If the factor-of-thirty improvements of the twisted molecules can be similarly made into a bulk material, one could expect electro-optic coefficients in excess of $3,000 \, pm/V$ - turning the promise of dye-doped polymers into a reality.

\vspace{1em}

\section{Conclusion}

We propose that the scale-invariant intrinsic hyperpolarizability is the best metric for comparing molecules.  We have shown that the off-resonance intrinsic hyperpolarizability can vary by more than an order of magnitude when compared with the often-used measure of $\beta_0$, showing that this is not an ideal method for comparison.  Furthermore, we have argued that for specific applications, it makes more sense to compare values of $\beta^{int}(\omega, \omega)$ at potential device wavelengths.  Analyzing the very best molecules, we show that if the new methods of poling used by Dalton and Jen and coworkers\cite{ma01.01,sulli07.01} are combined with the new paradigm of twisted chromophores of Marks and Ratner,\cite{Kang07.01} then it is possible to make electrooptic materials with electro-optic coefficients of several thousand pico-meters per volt.

{\bf Acknowledgements: } We thank the National Science Foundation
(ECS-0354736) and Wright Paterson Air Force Base for generously
supporting this work.

%\bibliography{\bibs}
\bibliographystyle{osajnl}

\clearpage

\end{document}